\documentclass[11pt]{article}

\usepackage{amssymb}
\usepackage{amsthm}
\usepackage{amsmath}

\usepackage{graphicx} 
\usepackage{psfrag}
\usepackage{subfigure}

\newcommand{\proofend}{\hspace*{\fill}\rule{0.2cm}{0.2cm}}

\newcommand{\textfrac}[2]{{\textstyle{\frac{#1}{#2}}}}

\newcommand{\calE}{{\cal E}}

\theoremstyle{plain}
\newtheorem{theorem}{Theorem}[section]

\newtheorem{lemma}[theorem]{Lemma}
\newtheorem{proposition}[theorem]{Proposition}

\theoremstyle{remark}

\begin{document}

\title{\sc Theory of Newtonian self-gravitating stationary spherically 
symmetric systems}

\author{\sc 
J.\ Mark Heinzle$^{1}$\thanks{Electronic address: {\tt
Mark.Heinzle@aei.mpg.de}}\ ,\ 
Alan D. Rendall$^{1}$\thanks{Electronic address: {\tt
Alan.Rendall@aei.mpg.de}}\ ,\
\ and
Claes Uggla$^{2}$\thanks{Electronic address:
{\tt Claes.Uggla@kau.se}}\\
$^{1}${\small\em Max-Planck-Institut f\"ur Gravitationsphysik, 
Am M\"uhlenberg 1}\\
{\small\em D-14476 Golm, Germany}\\
$^{2}${\small\em Department of Physics, University of Karlstad}\\
{\small\em S-651 88 Karlstad, Sweden}}

\date{}
\maketitle
\begin{abstract}

We investigate spherically symmetric equilibrium states of the
Vlasov-Poisson system, relevant in galactic dynamics.
We recast the equations into a regular three-dimensional system
of autonomous first order ordinary differential equations 
on a region with compact closure.
Based on a dynamical systems analysis we derive
theorems that guarantee that the steady state solutions
have finite mass and compact support.

\end{abstract}
\vfill
\newpage

\section{Introduction}
\label{introduction}

In this paper we investigate spherically symmetric equilibrium states of 
the  Vlasov-Poisson system. These equations describe a
collisionless gas of particles that interact by the Newtonian
gravitational field they generate collectively. Examples of systems
that can be modelled by these equations are ensembles of stars in
e.g., a galaxy, a globular cluster or an ensemble of galaxies in a
rich cluster~\cite{binney}.

A collisionless gas is characterized by the phase space mass density 
distribution
$f(t,{\bf x},{\bf v})$, where ${\bf x},{\bf v} \in \mathbb{R}^3$ and
where $f\,d^3{\bf x}\,d^3{\bf v}$ is the mass contained in 
the phase space volume element $d^3{\bf x}\,d^3{\bf v}$ at $({\bf x},{\bf v})$. 
The spatial mass density is given by 
\begin{equation}\label{density}
\rho({\bf x}) = \int\, f({\bf x},{\bf v})d^3{\bf v}\:.
\end{equation}
The Newtonian potential $V$ satisfies the Poisson equation
$\Delta V=4\pi\rho$. Under the assumption that $V$ tends to zero
at infinity and with suitable fall-off and regularity conditions on the 
function $\rho$ there is a unique potential $V<0$ for a given density $\rho$. It 
is assumed in the following that the function $f$ is non-negative and 
spherically symmetric, i.e., that it is invariant under simultaneous 
rotations of ${\bf x}$ and ${\bf v}$. The symmetry of $f$ is inherited by 
$\rho$ and $V$. Under these conditions the Poisson equation can be written 
as
\begin{equation}\label{poisson}
\frac{1}{r^2}(r^2 V^\prime)^\prime = 4\pi\rho(r)\:,
\end{equation}
where $r=|{\bf x}|$, ${}^\prime=d/dr$, and where the gravitational constant has been set to $1$. 
As the distribution function of a collisionless gas the function $f$
satisfies the Vlasov equation. The following is concerned with
time-independent solutions and in that case the Vlasov equation takes
the form
\begin{equation}\label{vlasov}
{\bf v}\cdot \frac{\partial f}{\partial 
{\bf x}}+\nabla V\cdot \frac{\partial f}{\partial {\bf v}}=0\:.
\end{equation}
Equations (\ref{density}), (\ref{poisson}) and (\ref{vlasov})
constitute the Vlasov-Poisson system in the time-independent, spherically
symmetric case. A definition which is useful when comparing models with
collisionless matter and models in which the matter is described by a fluid 
is the radial pressure defined by
\begin{equation}\label{pressure}
p_{\rm rad}({\bf x}) = \int\, v_r^2 f({\bf x},{\bf v})d^3{\bf v}\:,
\end{equation}
where $v_r = | \mathbf{v}\cdot\mathbf{x}|/r$ is the radial velocity of a particle.

For a time-independent spherically symmetric solution of the Vlasov-Poisson
system the distribution function depends only on the energy per 
unit mass and particle 
$E = \frac{1}{2}|{\bf v}|^2 + V(r)$ and the squared magnitude of the angular 
momentum $L^2= |{\bf x}\times {\bf v}|^2$. 
This statement was first proved in~\cite{batt} under appropriate regularity assumptions%
\footnote{In~\cite{batt} and~\cite{rr00} the function $f$ was interpreted as a number
  density, but this leads to the same mathematical problem as studied here.}%
;
it belongs to a class of statements known as Jeans theorems in the
astrophysical literature, cf.~\cite{binney}.
Conversely, any function of $E$ and 
$L^2$ satisfies the Vlasov equation. 

Now let $R \in (0,\infty]$ denote the radius of support of the system,
i.e., 
\begin{equation}
R=\{\inf r_1:\rho(r)=0\ {\rm for}\ r> r_1\}
\end{equation}
Because of physical applications, we are interested in
systems in equilibrium that have finite total mass. It follows 
from Theorem 2.1 of \cite{rr00} that the monotonically increasing 
potential satisfies $\lim_{r\rightarrow R} V(r) = V_R < \infty$
and that there is a cut-off energy $E_0=V_R$ such that  $f(E,L^2) = 0$ when 
$E \geq E_0$. We now introduce the definitions
\begin{equation}\label{binding}
\calE = E_0 - E\ , \quad \omega = E_0 - V = V_R - V\ ,
\end{equation}
where $\omega$ can be interpreted as a relative potential, and $\calE$ as the 
binding energy per unit mass for each particle. When 
$f(\calE,L^2)\neq 0$, then $\calE>0$, $\omega>0$, while $\calE$ and $\omega$ are 
zero at the boundary
$R$ of the system.

In terms of $\calE$ and $L^2$, the mass density takes the form:
\begin{equation}
\label{rhoofreta}
\rho (r,\omega) = \frac{2\pi}{r^2}\int^\omega_0\int^{L^2_{\rm max}}_0\,
f(\calE,L^2)|v_r|^{-1}dL^2d\calE\ ,
\end{equation}
where $L^2_{\rm max}=2r^2(E-V(r))=2r^2(\omega-\calE)$. 

The mass of the system inside the radius $r$ is determined by
\begin{equation}
\label{mprime}
m^\prime = 4\pi r^2\rho\ ,
\end{equation}
and thus $m= 4\pi\int_0^r s^2\rho ds$, assuming that there
is no central point mass.
As a consequence Poisson's equation takes the form
$(r^2 \omega^\prime)^\prime = m^\prime$, which leads to
\begin{equation}\label{etaprime}
\omega^\prime = -\frac{m}{r^2}\ .
\end{equation}
Hence $m$ and $V$ are monotonically increasing functions of $r$ while 
$\omega$ is monotonically decreasing.

Given a suitable choice of the dependence of $f$ on $\calE$ and $L^2$
the existence of corresponding global solutions with regular potential
can be shown \cite{batt}. 
What is much harder is 
to decide whether these solutions satisfy the condition that the total mass 
is finite. One approach to this, which will be pursued in the following, is 
to try to prove the stronger statement that the radius of the support is 
finite. In \cite{rr00} finiteness of the radius was proved for a large class
of solutions, which will be described in Section \ref{distf}. 
Many models used in the astrophysical literature are covered by the results of 
\cite{rr00}, but some are not. 
Motivated in part by this we consider a wider class of models in the following
and greatly extend the domain of validity of the results of \cite{rr00}.
The main assumptions used in 
\cite{rr00} concerned the behaviour of $f$ near $\calE=0$.
It turns out that if more general classes of solutions are to be handled 
it is often necessary to make assumptions concerning high values of $\calE$ 
as well.

The mathematical problem to be solved is to obtain information on
the qualitative behaviour of solutions of a system of ordinary 
differential equations. The technique which allows us to go beyond 
what was done previously is to reformulate the problem 
using new variables, since this enables us to apply the theory of 
dynamical systems to the resulting equations. The geometrical 
intuition resulting from the dynamical systems formulation played 
an important role in the development of the proofs. The new
variables are obtained by analogy with an approach of Heinzle
and Uggla \cite{hu1} to the Euler-Poisson system describing a
self-gravitating fluid.

The paper is organized as follows:
in Section~\ref{distf} we define and describe the class of distribution functions
we study in this article.
For these distribution functions, in Section~\ref{dyn}, 
we reformulate the static Vlasov-Poisson system
as a three-dimensional system of autonomous differential equations
on a state space with compact closure.
The system is subsequently analysed in Section~\ref{dynamicalsystemsanalysis}
by using methods from the theory of dynamical systems; in particular,
functions that are monotone along solutions of the dynamical system
play a key role.
Based on the results of Section~\ref{dynamicalsystemsanalysis}, 
the main theorems are stated and proved in Section~\ref{finitenessofradius}:
we formulate conditions that guarantee finiteness of the radius $R$ of
solutions. We conclude the paper in Section~\ref{outlook} with some examples 
and remarks, 
and give an outlook on further applications of the techniques developed here.

\section{Distribution functions}
\label{distf}

There exists a type of distribution functions that can be said to be among
the mathematically simplest, the ``generalized  
polytropes''
\begin{equation}\label{generalizedpolytropes} 
f(\calE,L^2) = 
\left\{ \begin{array}{ll}
\phi_- \calE^{n-3/2}L^{2l} & (\calE > 0) \\
0 & (\calE \leq 0) 
\end{array}\right.
\quad,
\end{equation}
where $n$, $l$ and $\phi_-$ are constants with $n>1/2$ and $l>-1$. 
Under these conditions the mass density of these models is well-defined, 
$\rho = \rho_- \,r^{2l}\,\omega^{n+l}$, where 
$\rho_- = 2^{l+3/2}\pi^{3/2} \Gamma(l+1)\Gamma(n-1/2)\Gamma(n+l+1)^{-1}$.
The models~(\ref{generalizedpolytropes}) exhibit invariance under 
scale transformations (of space and time),
which corresponds to
a symmetry of the Vlasov-Poisson equation; this leads to
mathematical simplification.
In particular, scale invariance gives rise to so-called 
homo\-logy invariants (e.g., \cite{hopf}, \cite{kippenhahn}, 
\cite{hu1}), i.e., quantities that are invariant under 
scalings, which are adapted to
the symmetries of the problem. 
The use of homology invariants turns out to be crucial for our considerations.

The mathematical analysis of static 
spherically symmetric solutions in the generalized polytropic case 
reduces to that of 
an ordinary differential equation known as the Emden-Fowler
equation. This equation has a long history intertwining 
mathematics and astrophysics which we have not attempted to 
reconstruct. We note only that Fowler \cite{fowler} introduced
this equation as a mathematical generalization of the Emden 
equation, the case $l=0$, whose importance in the theory of
stellar structure was well-known. Further early references can be
found in~\cite{fowler} and~\cite{batt}. It appears that 
the Emden-Fowler equation was introduced independently much later by 
H\'enon \cite{henon}, who derived it from a distribution function
of the form introduced above and coined the term \lq generalized 
polytropes\rq.

In this paper we discuss distribution functions that naturally generalize the polytropes:
\begin{equation}\label{fphi} 
f(\calE,L^2) = 
\left\{ \begin{array}{ll}
\phi(\calE)\,L^{2l} & (\calE > 0) \\
0 & (\calE \leq 0) 
\end{array}\right.
\quad,
\end{equation}
where $\phi(\calE)$ is a non-negative function which is measurable,
bounded on compact subsets of the interval $(0,\infty)$,
and integrable on $[0,1]$. Then, provided that 
$l>-1$, distribution functions of the type~(\ref{fphi}) give rise to a 
mass density
\begin{equation}\label{rhosplit}
\rho(r,\omega) =  C_l\,r^{2l}\,g_{l+1/2}(\omega)\: ,
\end{equation}
where $g_m$ ($m>-1$) is defined as
\begin{equation}\label{g}
g_m(\omega) := \int_0^{\omega}\, \phi(\calE)(\omega - \calE)^m\,d\calE\ ,
\end{equation}
and where $C_l:=2^{l+3/2}\pi^{3/2}\Gamma(l+1)/\Gamma(l+3/2)$. 
The radial pressure of the models~(\ref{fphi}) is given by 
$p_{rad}(r,\omega) = C_l r^{2l} g_{l+3/2}(\omega)/(l+3/2)$.

It is natural to define a polytropic index function $n(\omega)$ 
according to
\begin{equation}
n(\omega) =  -l + \frac{d\log g_{l+1/2}}{d\log\omega}\, \; .
\end{equation}
For the special case of the generalized polytrope
$\phi(\calE)\propto \calE^{n-3/2}$ the function $g_{l+1/2}(\omega)$ is 
given by $g_{l+1/2}(\omega) \propto \omega^{n+l}$, and therefore
$n(\omega)$ becomes a constant, $n(\omega) \equiv n$.

We call distributions that lead to a (bounded) function $n(\omega)$ and
satisfy $n(\omega) \rightarrow n_0$ for $\omega\rightarrow 0$ 
asymptotically polytropic in the low $\omega$ regime.
Distribution functions of this type with $n_0<3+l$ were treated in \cite{rr00}.
One of the main aims of the present work is to
treat more general distribution functions since this is needed for
applications.

If it is assumed in addition that $\phi$ 
satisfies
$\phi(\calE) \leq \mathrm{const}\: \calE^k$ for some $k > -1$ on a neighbourhood
of $\calE = 0$, then $g_{l+1/2}(\omega)$ is
continuous for $\omega >0$. Moreover, when $l > -1/2$,
$g_{l+1/2}(\omega) \in \mathcal{C}^1(0,\infty)$, and thus $n(\omega)$ is
continuous for all $\omega>0$; compare with the results of~\cite{rr00}.
For later purposes we require $n(\omega)$ to be of class $C^1$.
In order to achieve this we assume that $\phi$ is $C^1(0,\infty)$
such that $\phi^\prime(\calE)$ is bounded on compact subsets
of $(0,\infty)$ and
$\phi^\prime(\calE) \leq \mathrm{const}\: \calE^{k^\prime}$ for some $k^\prime > -2$
on a neighbourhood of $\calE = 0$.
Then it can be shown that $g_{l+1/2}(\omega)$ is $C^2$ for $\omega > 0$
and hence that $n(\omega)$ is $C^1$, provided that $l\geq -1/2$.
For the proofs, and a discussion of the case $-1 < l < -1/2$, see Appendix~\ref{diffassumptions}.
The regularity conditions on $\phi$ stated above will be assumed from now on.

\section{Dynamical systems formulation}
\label{dyn}

The main idea of this work is to reformulate the static Vlasov-Poisson system,
\begin{subequations}\label{vpsys}
\begin{align}
\frac{d m}{d r} & = 4 \pi r^2 \rho(r,\omega) \\
\frac{d \omega}{d r} & = -r^{-2} m \:,
\end{align}
\end{subequations}
cf.~(\ref{mprime}),~(\ref{etaprime}), as a three-dimensional system of
autonomous first order ordinary differential equations on a region 
with compact closure.

The system~(\ref{vpsys}) becomes an autonomous system of equations
by regarding $r$ as a supplementary dependent variable and 
introducing a new independent variable $\xi(r)$.
As the next step we perform a transformation of variables from 
$(m>0, r>0)$ to two dimensionless variables
\begin{equation} \label{uqomega} 
u= \frac{4\pi r^3\rho (r,\omega)}{m}\ ,\qquad 
q= \frac{m}{r \omega}\ ,\qquad .
\end{equation} 
Note that $uq=4\pi C_lr^{2+2l}g_{l+1/2}(\omega)/\omega$. Using the fact that 
$l>-1$ it is possible to compute $r$ in terms of $u$ and $q$. Then 
$m$ can be computed in terms of those variables and so the transformation 
is invertible for $\omega >0$.

Introducing $\xi := \log r$ as a new independent variable and converting the 
system~(\ref{vpsys}) 
to the new variables we obtain
\begin{subequations}\label{uqomegaeq}
\begin{align}
\label{ueq}
&\frac{du}{d\xi} \,=\, u \,\big( 3 - u + 2l - n(\omega) q- l q \big)\ ,\\
\label{qeq}
&\frac{dq}{d\xi} \,=\,q \,\big( -1 + u + q  \big)\ ,\\
\label{omegaeq}
&\frac{d\omega}{d\xi} \,= \,-q \,\omega \ , 
\end{align}
\end{subequations}

We proceed by defining bounded variables in order to obtain 
a dynamical system on a bounded state space.
Recall that $u,q,\omega>0$ and define
bounded variables $U,Q,\Omega$, $(U,Q,\Omega) \in (0,1)^3$, by
\begin{equation}\label{boundedvar}
U = \frac{u}{1+u}\ , \quad
Q = \frac{q}{1+q}\ , \quad
\Omega = \frac{\omega}{1+\omega} \quad .
\end{equation}
If a solution $u(\xi), q(\xi),\omega(\xi)$ is given, a new independent 
variable $\lambda$ can be introduced by the relation 
$d\lambda/d\xi = (1 - U)^{-1}(1 - Q)^{-1}$. This yields a solution of the 
system of equations
\begin{subequations}\label{UQOmega}
\begin{align}
\label{Ueq}
\frac{dU}{d\lambda} &=  U(1-U)\,\big[(1-Q)\big(3+2l - (4+2l) U\big) - 
(n(\Omega)+l)\,Q(1-U)\big] \\
\label{Qeq}
  \frac{dQ}{d\lambda} &= 
  Q(1-Q)\big[(2U-1)(1-Q) + Q (1-U)\big] \\
\label{Omegaeq}
  \frac{d\Omega}{d\lambda} &= -\Omega(1-\Omega)Q(1-U) \: ,
\end{align}
\end{subequations}
where $n(\Omega)$ is $n(\omega)|_{\omega(\Omega)}$.

The r.h.s.\ of the system~(\ref{UQOmega}) is $\mathcal{C}^1$ in $\Omega$, 
because of the previous assumptions
on the distribution function, and polynomial in $U$ and $Q$.
Hence,
it is natural to smoothly extend the system to the side faces of the 
cube, so that the system of equations~(\ref{UQOmega}) forms a $\mathcal{C}^1$ dynamical system on 
the state space $[0,1]^2 \times (0,1)$.
We observe that the side faces are invariant subspaces of the system,
moreover, they contain the attracting sets for any orbit in the state space,
as discussed in the next section. Note that a 
solution of (\ref{UQOmega}) which is global in $\lambda$ need neither
correspond to a solution of (\ref{uqomegaeq}) which is global in $\xi$,
nor to a solution of (\ref{vpsys}) which is global in $r>0$.

For generalized polytropes $n(\Omega)\equiv \mathrm{const}$, and 
the equation for $\Omega$ decouples from the equations for $U$ and $Q$.
This is a direct consequence of the fact that the variables $U$, $Q$ are
homology invariants. However, in the case of a general distribution function,
$n(\Omega)$ gives rise to the three-dimensional coupled system~(\ref{UQOmega}).

\section{Dynamical systems analysis}
\label{dynamicalsystemsanalysis}

\begin{figure}[htp]
        \psfrag{L1}[cc][bc]{{\small $L_1$}}
        \psfrag{L2}[cc][bc]{{\small $L_2$}}
        \psfrag{L3}[cc][bc]{{\small $L_3$}}
        \psfrag{L4}[cc][bc]{{\small $L_4$}}
        \psfrag{O}[cc][cc]{{$\Omega$}}
        \psfrag{Q}[cc][cc]{{$Q$}}
        \psfrag{U}[cc][cc]{{$U$}}
        \centering
        \includegraphics[width=0.4\textwidth]{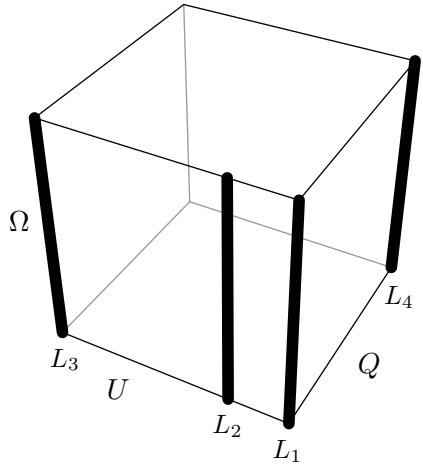}
        \caption{The state space $[0,1]^2\times(0,1)$ with fixed points.
                There exist four ``fixed lines'', i.e., 
                lines consisting of fixed points ($L_1,L_2, L_3, L_4$);
                the position of $L_2$ depends on $l$.}
        \label{cube_sides}
\end{figure}

Consider the dynamical system~(\ref{UQOmega}) on the state space 
$[0,1]^2\times(0,1)$.
The side faces of the cube are invariant subspaces, where the fixed points of the 
system are located, cf. Figure~\ref{cube_sides}. Table~\ref{tab:UQcube_sides} 
lists
the fixed points together with the eigenvalues of the linearizations of the 
system at the fixed points. It is elementary to prove the facts summarized
in this table. We observe that the flows on the side faces $U=0$, $U=1$, and 
$Q=0$ possess a simple structure. In each of these cases the evolution for the 
variable other than $\Omega$ does not contain $\Omega$ and the induced system
does not depend on $n(\Omega)$.

\begin{table}[ht]
  \begin{center}
    \begin{tabular}{|c|cccc|c|}
      \hline
      Fixed point & $U$ & $Q$ & $\Omega$ &   & Eigenvalues\\  \hline 
        & & & & &  \\[-1ex]
      $L_1$ & 1 & 0 & $\Omega_0$ &   & $1\ , \ 1\ , \ 0$ \\[0.5ex]
      $L_2$ & $\textfrac{3+2l}{4+2l}$ & 0 & $\Omega_0$ &   &
      $-\textfrac{3+2l}{4+2l}\ , \ \textfrac{1+l}{2+l}\ , \ 0$\\[0.5ex]
      $L_3$ & 0 & 0 & $\Omega_0$ &   & $3+2l\ , \ -1\ , \ 0$ \\[0.5ex]
      $L_4$ & 1 & 1 & $\Omega_0$ &   & $0 \ , \ 0 \ , \ 0$ \\ \hline
    \end{tabular}
  \end{center}
    \caption{Local properties of the fixed points on the side faces.
      Here $\Omega_0$ is a parameter belonging to the interval $(0,1)$ in each
      case.}
    \label{tab:UQcube_sides}
\end{table}

Since the coefficients of the system are $C^1$ we can conclude that for
any initial data there exists a unique local solution and that it extends 
to a global solution provided $\Omega$ remains bounded away from zero and
one on any finite interval. 
The latter condition always holds since $d/d\lambda (\log\Omega)$ and
$d/d\lambda (\log (1-\Omega))$ are bounded along any solution. 

An important tool for the local analysis of dynamical systems near
a stationary point is the reduction theorem (see \cite{kirchgraber}, p. 48). 
It says that the flow of any $C^1$ dynamical system is topologically 
equivalent in a neighbourhood of a stationary point to a product of a 
standard saddle with the flow on any centre manifold.

The lines $L_2$ and $L_3$ are transversely hyperbolic saddles in the
following sense. The linearization of the system about any point of
one of these lines has a zero eigenvalue whose corresponding 
eigendirection points along the line. The other two eigenvalues are 
non-zero, real and have opposite signs. Each of these lines is a centre 
manifold for any of its points. By the reduction theorem this
means that the flow near one of these points looks topologically like
the product of a hyperbolic saddle with a line. 
The stable manifolds of points on $L_2$ and the stable and 
unstable manifolds of points on $L_3$ lie in the boundary planes. It can
be concluded from these facts that
the boundary planes involved are mapped onto coordinate planes by the mapping 
which does the reduction. As a consequence any interior solution which
has an $\omega$-limit point on $L_2$ also has $\omega$-limit points 
with $Q=0$ not lying on $L_2$ and $\omega$-limit points in the interior.
Any interior solution which has an $\omega$-limit point ($\alpha$-limit point) 
on $L_3$ has $\omega$-limit points ($\alpha$-limit points) 
with $Q=0$ not lying on $L_3$ and $\omega$-limit points ($\alpha$-limit points) 
with $U=0$ not lying on $L_3$.
We will show below 
in Proposition~\ref{globalomega} that these
properties of the transversely hyperbolic saddles in combination with
the monotonicity properties of the system~(\ref{UQOmega}) imply that
an interior solution cannot have an $\omega$-limit point on $L_2$, 
and neither an $\alpha$- nor an $\omega$-limit point
on $L_3$.

Similarly, $L_1$ is a transversely hyperbolic source,
where both non-zero eigenvalues of the linearization are positive and $L_1$
is a centre manifold for any of its points. It follows from the reduction 
theorem that any point of $L_1$ has a neighbourhood such that every interior 
solution which intersects that neighbourhood converges to a point of $L_1$ as 
$\lambda\to -\infty$ and no interior solution enters that neighbourhood
from outside. 

Another important tool which we will use in the following is the 
monotonicity principle \cite{wainwright}, which will now be stated.
Let $\phi_t$ be the flow of a dynamical system on an open set 
$U\subset \mathbb{R}^n$ 
with $S\subset U$ an invariant set. Let $Z$ be a $C^1$ 
function on $S$ 
whose range is the interval $(a,b)$ where $a\in\mathbb{R}\cup\{-\infty\}$ and 
$b\in\mathbb{R}\cup\{+\infty\}$ and $a<b$. If $Z$ is decreasing on orbits
in $S$, then for all ${\bf x}\in S$ the $\alpha$- and $\omega$-limit sets
of the orbit starting at $x$ are disjoint from $S$. Moreover it
cannot be the case that $Z$ tends to $b$ at any $\omega$-limit point 
or that $Z$ tends to $a$ at any $\alpha$-limit point. In 
\cite{wainwright} this was stated only for $U=\mathbb{R}^n$ but
the same proof works for general $U$.

\begin{proposition}\label{globalomega} \textbf{(Global dynamics)}.
The $\omega$-limit of every interior orbit is located on $\Omega=0$.
\end{proposition}

\proof
The proof of the theorem is based on the monotonicity principle.
The function $\Omega$ is a strictly monotonically decreasing function on 
$(0,1)^3$ and on the side faces $U=0$ ($Q\ne 0$) and $Q=1$ ($U\ne 1$). Accordingly, 
the monotonicity principle yields that the $\omega$-limit of an interior 
orbit must be located on $U=1$, $Q=0$ or $\Omega=0$. Suppose that an interior  
solution has an $\omega$-limit point $p$ with $U=1$, $Q>0$ and $\Omega>0$. 
There is a solution on the boundary $U=1$ passing through $p$. Because of 
the simple structure of the dynamical system on the surface $U=1$ it follows 
that the new solution has an $\omega$-limit point with $Q=1$ and $\Omega>0$. 
This implies that the original solution also has a limit point with $U=1$,
$Q=1$ and $\Omega>0$. This limit point is on $L_4$ and it will be shown 
in Lemma \ref{L4Lemma} that this is not possible. Suppose next that an
interior solution has an $\omega$-limit point $p$ with $Q=0$, $0<U<1$ and 
$\Omega>0$. There is a solution on the boundary $Q=0$ passing through $p$. 
Because of the simple structure of the dynamical system on the surface $Q=0$ 
it follows that the new solution has an $\omega$-limit point on $L_2$ with 
$\Omega>0$. Hence the original solution also has an $\omega$-limit point with 
these properties. Because of the structure of $L_2$ as a tranversely hyperbolic
saddle this implies that the original solution has an interior 
$\omega$-limit point, in contradiction to what has already been proved. 
Suppose that an interior solution has an $\omega$-limit point on $L_3$. 
Since $L_3$ is a transversally hyperbolic saddle, 
the solution must also possess an $\omega$-limit point $p$ with $Q=0$, $0<U<1$;
this case has already been excluded above.
Finally, if an interior solution has an $\omega$-limit point on $L_1$, 
then the fact that $L_1$ is a transversely hyperbolic source also leads to a 
contradiction. Thus $U=1$ and $Q=0$ cannot contain an $\omega$-limit of an 
interior orbit and this leaves $\Omega=0$ as the only attracting set.
\proofend

The fixed points on $L_4$ constitute a special case because they
are not only non-hyperbolic, but the linearization of the 
dynamical system~(\ref{UQOmega}) at such a fixed point has three zero 
eigenvalues.

\begin{lemma}\label{L4Lemma}
No interior orbit has an $\alpha$- or $\omega$-limit point on $L_4$.
\end{lemma}

\proof
We rewrite the system~(\ref{UQOmega}) (where $U<1$, $Q<1$)
in cylindrical coordinates centred at $(1,1,0)$. Define
\begin{equation}
\tilde{r} = \sqrt{(1-U)^2+(1-Q)^2} \qquad
\phi = \arctan \frac{1-Q}{1-U} \qquad
z = \Omega\:.
\end{equation}
The coordinate $\phi$ ranges in $(0,\pi/2)$, $\tilde{r}$ is in $(0,\tilde{r}_{\max})$,
where $0<\tilde{r}_{\max}<1$ can be chosen arbitrarily.
We introduce the new independent variable
$\tilde{\lambda}$ via $d\tilde{\lambda}/d\lambda = \tilde{r}$.
By this the system~(\ref{UQOmega}) is transformed to
an equivalent dynamical system in $(\tilde{r},\phi,z)$,
\begin{equation}\label{L4dynsys}
\frac{d\tilde{r}}{d\tilde{\lambda}} = f_r(\tilde{r},\phi,z) \quad
\frac{d\phi}{d\tilde{\lambda}} = f_\phi(\tilde{r},\phi,z) \quad
\frac{dz}{d\tilde{\lambda}} = f_z(\tilde{r},\phi,z) \:.
\end{equation}
The $f_i(\tilde{r},\phi,z)$ are of the form
\begin{subequations}
\begin{align}
f_r(\tilde{r},\phi,z) & = f_{r,1}(\phi,z) \tilde{r} + f_{r,2}(\phi,z) \tilde{r}^2 + f_{r,3}(\phi,z) 
\tilde{r}^3 \\
f_\phi(\tilde{r},\phi,z) & = f_{\phi,0}(\phi,z) + f_{\phi,1}(\phi,z) \tilde{r} 
+ f_{\phi,2}(\phi,z) \tilde{r}^2 \\
f_z(\tilde{r},\phi,z) & = f_{z,0}(\phi,z)  + f_{z,1}(\phi,z) \tilde{r} \ ,
\end{align}
\end{subequations}
where the $f_{i,j}$ are polynomials in $\cos\phi$, $\sin\phi$.
Thus the system can be smoothly extended to also include the boundaries
$\tilde{r}=0$, $\phi=0$, and $\phi=\pi/2$, so that the state space is
$[0,\tilde{r}_{\max})\times[0,\pi/2]\times (0,1)$.
From the construction it follows that
the subset $\tilde{r}=0$ can be regarded
as a blow-up of the fixed line $L_4$; clearly, $\tilde{r}=0$ is an invariant subset.

A fixed point analysis of the system reveals that there exist only 
the fixed points $(0,\pi/2,z_0)$ ($z_0\in (0,1)$) on $\tilde{r}=0$; elsewhere
$z$ is monotone. These fixed points turn out to be transversely 
hyperbolic saddle points and hence no orbit with $\tilde{r}>0$ can have an $\alpha$- 
or $\omega$-limit point with $\tilde{r}=0$. This establishes the claim of the lemma.
\proofend

\begin{proposition}\label{globalomega2}
Assume that $n(\Omega) \leq 3 + l$ for all $\Omega \leq \Omega_0$
(for some arbitrary $0<\Omega_0<1$). Then the $\omega$-limit of every 
interior orbit
lies on 
$Q=1$, $\Omega=0$.
If $-l+\epsilon \leq n(\Omega) \leq 3 + l$ for all $\Omega \leq \Omega_0$ 
(for some $\epsilon>0$)
then the $\omega$-limit of every interior orbit is the point $(0,1,0)$;
if $n(\Omega) \leq -l-\epsilon$ for all $\Omega \leq \Omega_0$ 
(for some $\epsilon>0$), then the $\omega$-limit is $(1,1,0)$.
\end{proposition}

\proof
Consider the function 
\begin{equation}\label{Z1II}
Z = \left(\frac{U}{1-U}\right)\left(\frac{Q}{1-Q}\right)^{(3+2l)} \:.
\end{equation}
The function $Z$ is strictly monotonically increasing on all interior 
orbits, 
\begin{equation}\label{dZ1II}
\frac{dZ}{d\lambda} = (2 (l+1) U (1-Q) + (3+l-n) Q (1-U))Z > 0 \ .
\end{equation}
Since $d\log Z/d\lambda >0$ it follows that $Z$ tends to a limit,
finite or infinite as $\lambda\to\infty$. In the latter case, 
$Z\rightarrow \infty$, we observe $(1-U)(1-Q)\rightarrow 0$,
so that the $\omega$-limit of the orbit must lie on $(U=1) \cup (Q=1)$.
Now assume the first case, $\lim_{\lambda\rightarrow\infty}\log Z < \infty$.
From~(\ref{dZ1II}) we obtain
$d\log Z/d\lambda \geq 2(l+1) U (1-Q)$, hence
\begin{equation}
\int^\infty U (1-Q) d\lambda < \infty\:.
\end{equation}
Since, firstly, $U(1-Q) > 0$ and, secondly, the derivative of $U (1-Q)$ is 
always bounded,
we conclude that $U (1-Q) \rightarrow 0$ as $\lambda\rightarrow\infty$.
Assume that $Q\not\rightarrow 1$. Then there is sequence 
$\lambda_n$ such that
$1-Q(\lambda_n) \geq \mathrm{const} >0$ and $U(\lambda_n)\rightarrow 0$ for 
$n\rightarrow \infty$. This implies that $Z(\lambda_n)\rightarrow 0$, a 
contradiction. Thus, $Q\rightarrow 1$ in
the limit $\lambda\rightarrow \infty$, and the $\omega$-limit of the orbit 
must lie on $Q=1$.

In combination with Proposition~\ref{globalomega}
this yields that the $\omega$-limit of every interior orbit must
lie on $(U=1) \cap (\Omega=0)$ or $(Q=1) \cap (\Omega=0)$.

\begin{figure}[htp]
        \psfrag{L1}{{$L_1$}}
        \psfrag{L2}{{$L_2$}}
        \psfrag{L3}{{$L_3$}}
        \psfrag{L4}{{$L_4$}}
        \psfrag{Q}[cb][Bl]{{$\begin{array}{c}\uparrow\\ Q\end{array}$}}
        \psfrag{U}{{$U\rightarrow$}}
        \centering
        \subfigure[]{
                \label{S_1}
                \includegraphics[height=0.4\textwidth]{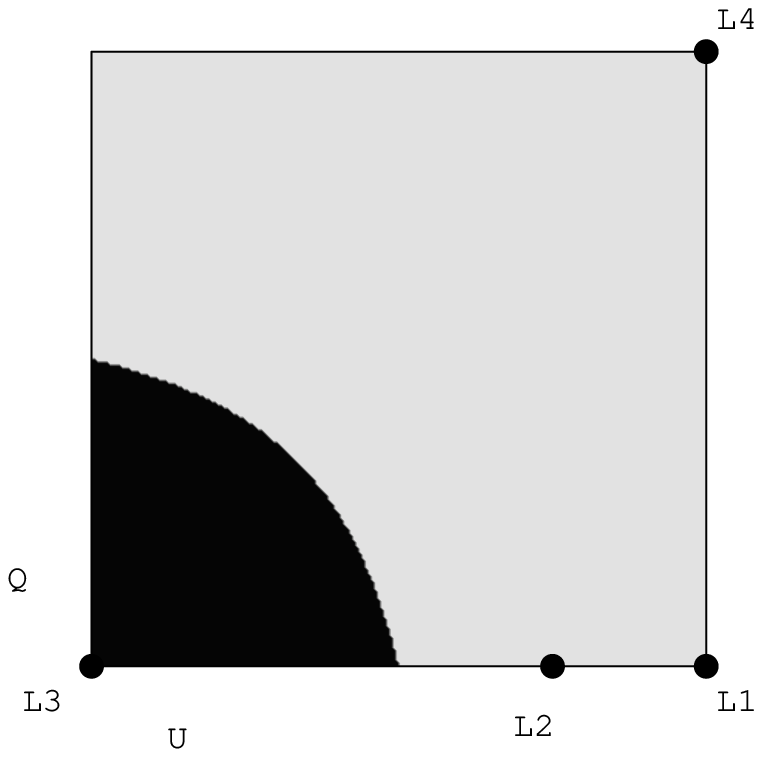}}\qquad
        \subfigure[]{
                \label{S_3}
                \includegraphics[height=0.4\textwidth]{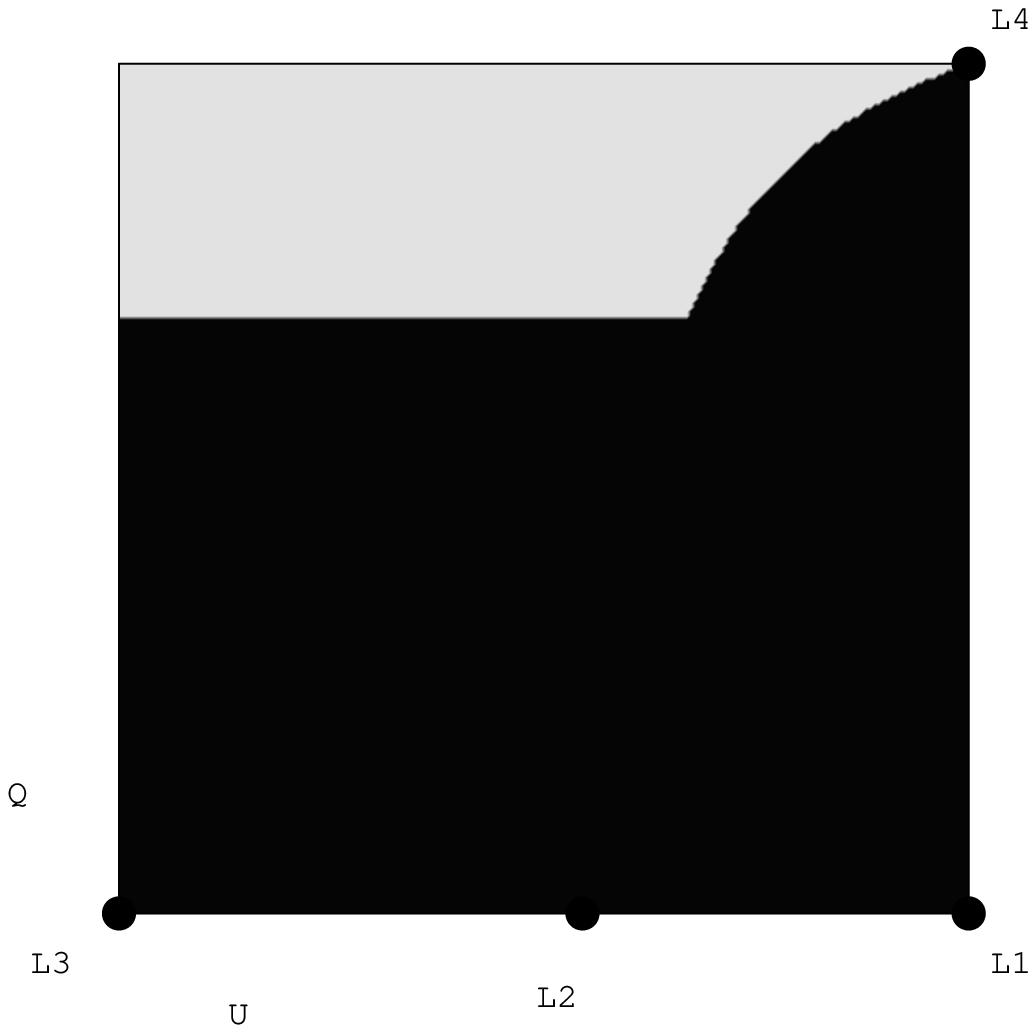}}
        \caption{Subfigure~(a) shows the invariant subset $S_1$ --- the 
	  light-colored region.
                 In Subfigure~(b) the set $S_3$ is depicted for value of $l$ 
		 close to $-1$.}
\end{figure}

Consider the
set $S_1=\{(U,Q,\Omega)\:|\:(2U-1)(1-Q) + Q (1-U) > 0\}$ in the state space, 
see Figure~\ref{S_1}.
Note that $U=1$ and $Q=1$ are contained in $S_1$.
It is not difficult to show that $S_1$ is a future invariant set:
firstly, note that $dQ/d\lambda =0$ on the boundary of $S_1$, cf.~(\ref{Qeq}).
Secondly, we observe that 
\begin{equation}
\frac{d U}{d\lambda} = U(1-U)(1-Q) [3+l-n + 2 (n-2) U]  \geq 0 \quad 
\mbox{on}\quad \partial S_1\ ,
\end{equation}
so that $\partial S_1$ acts as a ``semipermeable membrane''.

On $S_1$ the function $Q$ is strictly monotonically increasing except
when $Q=0$ and $Q=1$. Application of the monotonicity principle yields that
the $\omega$-limit of every orbit must lie on $Q=1$. 
Hence the first part of the proposition is established.

Assume $-l +\epsilon < n(\Omega) \leq 3 + l$ on $(0,\Omega_0]$ (for some $\epsilon>0$). 
Consider the set 
$S_2 =\{(U,Q,\Omega)\:|\: Q > \max\left(\frac{1}{2},\sup_\Omega 
\frac{3+2l}{(3+2l) +(l+n)}\right)\}$
in $[0,1]^2\times(0,\Omega_0)$. 
$S_2$ is future invariant, which is because
$d\Omega/d\lambda \leq 0$ and 
$dQ/d\lambda \geq 0$ on $\partial S_2$. 
On $S_2$ the function $U$ is strictly monotonically decreasing except when 
$U=0$ or $U=1$. 
Accordingly, the monotonicity
principle yields that the $\omega$-limit of every orbit must lie on $U=0$.
Combining this result with the previous statement we see that 
the $\omega$-limit of every interior orbit is located
on \mbox{$(Q=1) \cap (\Omega=0) \cap (U=0)$}, i.e., it is the point $(0,1,0)$.

Assume $n(\Omega)+l < -\epsilon$ on $(0,\Omega_0]$ (for some $\epsilon>0$)
and consider the set
$S_3$ depicted in Subfigure~\ref{S_3}. Its boundaries are $Q=1 - \delta$ (for some
appropriate $\delta$) and the
surface given by 
\begin{equation}
U = \frac{(3+2l)(1-Q)-\epsilon Q}{(3+2l)(1-Q)- \epsilon Q +(1-Q)}\ .
\end{equation}
$S_3$ can be shown to be future invariant, and $dU/d\lambda > 0$
on $S_3$. Using the monotonicity principle in combination
with the previous statement of the proposition, we see
that the $\omega$-limit on every orbit must be the point $(1,1,0)$, as claimed.
\proofend

\begin{proposition}\label{globalomega3}
Assume that $n(\Omega) \leq 5 + 3 l$ for all $\Omega \leq \Omega_0$
(for some $0<\Omega_0<1$) and $n(\Omega) \not\equiv 5 + 3 l$ in a 
neighbourhood of 
$\Omega =0$. 
Then the $\omega$-limit of every orbit originating from $L_1 
\cap (\Omega\leq \Omega_0)$ 
or $L_2\cap (\Omega\leq\Omega_0)$
lies on $Q=1$, $\Omega=0$.
If $-l+\epsilon \leq n(\Omega) \leq 5 + 3 l$ 
(for some $\epsilon>0$)
for all $\Omega$ small enough,
then the $\omega$-limit of every orbit is the point $(0,1,0)$.
\end{proposition}

\begin{figure}[htp]
        \psfrag{L1}{{$L_1$}}
        \psfrag{L2}{{$L_2$}}
        \psfrag{L3}{{$L_3$}}
        \psfrag{L4}{{$L_4$}}
        \psfrag{Q}[cb][Bl]{{$\begin{array}{c}\uparrow\\ Q\end{array}$}}
        \psfrag{U}{{$U\rightarrow$}}
        \centering
        \includegraphics[height=0.4\textwidth]{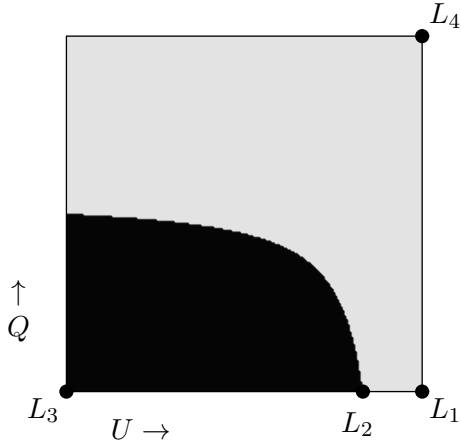}
        \caption{The future invariant set $\Psi>0$ for $l=1$.}
        \label{Psiregion}
\end{figure}

\proof
Consider the function 
\begin{equation}
\Phi = -\frac{1}{2} \left(\frac{U}{1-U}\right)^{\frac{1}{2(1+l)}} 
\left(\frac{Q}{1-Q} \right)^{\frac{3+2 l}{2(1+l)}} 
\:\left( 1- 
\frac{Q}{1-Q} - \frac{U}{(3+2 l)(1-U)}\right)
\end{equation}
defined for $(U,Q) \in [0,1)\times [0,1)$
and consider the surface $\Phi(U,Q)=0$ in the state space. In the case
$n(\Omega)\equiv 5+3 l$ the function $\Phi$ is a conserved quantity of the flow.
Under the given assumptions the set $(\Phi>0) \cap (\Omega\leq \Omega_0)$ 
is a future invariant subspace of the state space
because $(\Phi=0) \cap (\Omega\leq\Omega_0)$ is a semipermeable membrane: 
\begin{eqnarray}\label{Phider}
\nonumber
\lefteqn{\frac{d\Phi}{d\lambda} = \frac{d\Phi}{d\lambda}\:\Big|_{n(\Omega)} =
\frac{d\Phi}{d\lambda}\:\Big|_{n(\Omega)} - 
\frac{d\Phi}{d\lambda}\:\Big|_{n = 5+3l} = }\\[1ex]
\nonumber
& & = \frac{\partial\Phi}{\partial U} 
\left( \frac{d U}{d\lambda}\:\Big|_{n(\Omega)} - 
\frac{d U}{d\lambda}\:\Big|_{5+3l}\right) +
\frac{\partial\Phi}{\partial Q} 
\left( \frac{d Q}{d\lambda}\:\Big|_{n(\Omega)} - 
\frac{d Q}{d\lambda}\:\Big|_{5+3l}\right) = \\[1ex]
& & = \frac{\partial\Phi}{\partial U}\: U (1-U)^2 Q \,
\big[\,(5+3l)-n(\Omega)\,\big]\ .
\end{eqnarray}
Now,
\begin{equation}\label{dPhidU}
\frac{\partial\Phi}{\partial U} =
\frac{Q}{4}\, \left(\frac{U Q}{(1-U)(1-Q)}\right)^{\frac{1}{2(1+l)}}\,
\frac{(2 U -1)(1-Q) + Q (1-U)}{(1+l) (1-Q)^2 (1-U)^2 U}\:.
\end{equation}
From~(\ref{dPhidU}) we see that $\partial\Phi/\partial U > 0$ if and only if
$(U,Q) \in S_1$, cf. Fig.~\ref{S_1}. Since $\Phi = 0$ is contained 
in $S_1$, $\partial\Phi/\partial U > 0$ and therefore $d \Phi/d\lambda >0$, as claimed.
The surface $\Phi=0$ and the set $\Phi>0$ 
are depicted in Figure~\ref{Psiregion}.

Since the set $(\Phi>0)\cap (\Omega\leq \Omega_0)$ 
is future invariant, every orbit originating from $L_1 \cap 
(\Omega\leq\Omega_0)$ 
in entirely contained in this set. 
The same is true for orbits that originate from $L_2$ but this is more delicate 
to prove, since $L_2$ lies on the boundary of the set $\Phi>0$. 
Recall, first, that $\partial\Phi/\partial U$ is positive on $S_1$, which is a 
neighbourhood of the surface $(\Phi=0)\backslash \{(0,1/2,\Omega)\,| \,\Omega\in(0,\Omega_0]\}$. 
Now consider a solution that
starts from $L_2$ with $\Omega=\Omega_1\leq \Omega_0$.
Along the solution, $\Phi\rightarrow 0$ for $\lambda\rightarrow -\infty$ holds. 
The orbit is in $S_1$ as long as it is sufficiently close to $L_2$, i.e.,
for sufficiently large negative $\lambda$ we observe $\partial\Phi/\partial U>0$.
If $\Omega_1$ is such that $n(\Omega_1)<5+3l$, then 
$d\Phi/d\lambda > 0$ in the limit $\lambda\rightarrow -\infty$, 
hence $d\Phi/d\lambda > 0$
for all $\lambda$ in a neighbourhood of $-\infty$.
We conclude that $\Phi$ is positive for all large negative $\lambda$
and thus for all $\lambda$, since $\Phi>0$ is a future invariant set. 
If $n(\Omega_1)=5+3l$ but
$n(\Omega)<5+3l$ for any $\Omega<\Omega_1$ then the same argument applies.
If neither of these conditions are satisfied let $\Omega_2$ be the smallest
number for which $n(\Omega)=5+3l$ on the interval $[\Omega_2,\Omega_1]$
and let $\lambda_2$ be such that $\Omega(\lambda_2) = \Omega_2$.
For $\lambda\leq \lambda_2$ the solution lies on the surface $\Phi =0$; however,
$d\Phi/d\lambda>0$
for $\lambda$ slightly larger than $\lambda_2$, 
since the solution is in $S_1$. 
Thus in all cases $\Phi$ eventually becomes positive.

To continue the proof we proceed as in Proposition~\ref{globalomega2}.
Since $(\Phi>0) \subset S_1$, 
$Q$ is strictly monotonically increasing along interior solutions in 
$(\Phi>0)\cap (\Omega\leq \Omega_0)$;
by the monotonicity principle the $\omega$-limit of an interior
solution must lie on $Q=1$.
This proves the first assertion of the proposition. 

To conclude the proof, recall that $U$ is strictly monotonically
decreasing on $S_2$ when 
$n(\Omega) +l \geq \epsilon$ holds, which is the case for all small $\Omega$
by assumption.
The monotonicity principle yields that the $\omega$-limit
of an interior orbit is the point $(0,1,0)$. 
\proofend

\begin{proposition}\label{globalalpha} \textbf{(Global dynamics)}.
The $\alpha$-limit of an interior orbit is a fixed point on $L_1$, $L_2$,
or contained in $\Omega=1$.
\end{proposition}

\proof
The function $\Omega$ is a strictly monotonically decreasing function on 
$(0,1)^3$, and on the side faces $U=0$ ($Q\ne 0$) and $Q=1$ ($U\ne 1$). Accordingly, 
the monotonicity principle yields that the $\alpha$-limit of an interior orbit 
must be located on $U=1$, $Q=0$, or $\Omega=1$. 
The line $L_4$ cannot act as an $\alpha$-limit set for interior orbits,
see Lemma~\ref{L4Lemma}.  
Suppose that an interior solution has 
an $\alpha$-limit point $p$ with $U=1$, $0<Q<1$ and $\Omega<1$. Then 
the entire orbit through $p$ must be contained in the $\alpha$-limit set,
and hence a point on $L_4$ because of the structure of the flow; a contradiction
to what has already been shown.
Therefore any 
$\alpha$-limit point with $\Omega\neq 1$ satisfies $Q=0$. 
Suppose that an interior orbit has an $\alpha$-limit point on $L_3$.
Since $L_3$ is a transversally hyperbolic saddle, the orbit must also
possess an $\alpha$-limit point with $U=0$, $Q>0$, a contradiction.
On $Q=0$,  
$\alpha$-limit points with $(3+2 l)/(4+2 l) \neq U < 1$ are excluded, since
$L_1$ is a tranversely hyperbolic source and because of the fact that 
the transversely hyperbolic saddle $L_3$ cannot act as an $\alpha$-limit set as
shown above.
Therefore, any 
$\alpha$-limit points with $\Omega\ne 1$ must be on $L_1$ or $L_2$.
\proofend

If a solution of the Vlasov-Poisson system possesses a regular potential $V$, 
i.e., if $V(r)$ is $C^1$ on $[0,\infty)$,
then $\Omega$ converges to some limit $\Omega_0 \in (0,\infty)$
(and thus $\omega\rightarrow \omega_0$) as $r\rightarrow 0$,  whereby
$Q\rightarrow 0$, and $U\rightarrow (3+2 l)/(4+2 l)$ ($r\rightarrow 0$).
To show this we use~(\ref{uqomega}) and 
$g_{l+1/2}(\omega) = g_{l+1/2}(\omega_0) + o(1)$ as $r\rightarrow 0$.
The corresponding solution $(U,Q,\Omega)(\lambda)$ of 
(\ref{UQOmega}) exists globally to the past
and the above limits are attained as $\lambda\to -\infty$;
the solution is therefore associated with an orbit
whose $\alpha$-limit resides on $L_2$.
Although they have a regular potential, these solutions do not necessarily
possess a regular density function $\rho(r)$ since
$\rho \propto r^{2 l} g_{l+1/2}(\omega_0)$ as $r\rightarrow 0$, and therefore
$\rho\rightarrow \infty$ for $l<0$.
Nevertheless, for reasons of brevity, we refer to these solutions
as regular solutions.

\section{Finiteness of the radius of the support}
\label{finitenessofradius}

For the theorems, let us restate the regularity requirements on $\phi$:
we assume that $\phi(\calE)$ is $\mathcal{C}^1(0,\infty)$, and that 
$\phi$, $\phi^\prime$ are functions, which are bounded on
compact subsets of $(0,\infty)$. It is required that
$\phi(\calE) \leq \mathrm{const}\: \calE^k$ for some $k > -1$ 
and $\phi^\prime(\calE) \leq \mathrm{const}\: \calE^{k^\prime}$ for some $k^\prime > -2$
on a neighbourhood of $\calE = 0$.
When $l<-1/2$, $\calE \phi(\calE)$ 
must be H\"older-continuous with an
index strictly greater than $-l-1/2$.

\begin{theorem}\label{maintheorem1}
Let $\phi(\calE)$ be such that $n(\omega) \leq 3 + l$ for all $\omega \leq \omega_0$
(for an arbitrarily small $\omega_0>0$).  
Then every associated solution of the Vlasov-Poisson system has finite mass 
and radius.
\end{theorem}

\proof
We investigate the relation $dr/d\lambda =  r(1-U) (1-Q)$.
Integration yields
\begin{equation}\label{someint}
R \,=\, r_0 \exp\left[ \int\limits_{\lambda_0}^\infty (1-U(\lambda))
(1-Q(\lambda)) d\lambda \,\right]\ ,
\end{equation}
where $\lambda_0$, $r_0$ are such that $(U,Q,\Omega)(\lambda_0)$ corresponds to
$(m(r_0),\omega(r_0),r_0)$.
Thus, to prove that $R < \infty$, it suffices to show
that the integral in~(\ref{someint}) is finite.

From Proposition~\ref{globalomega2} we know that the $\omega$-limit of every 
interior orbit 
is a subset of the set \mbox{$(\Omega=0)\cap(Q=1)$}.
Define $\delta Q = 1- Q$ and $\delta U = 1 -U$. Since $Q$ is strictly 
monotonically increasing on the set $S_1$, 
see Figure~\ref{S_1}, $\delta Q$ is strictly monotonically decreasing on 
every orbit
for sufficiently large $\lambda$.
Written in $\delta Q$, $\delta U$, Eq.~(\ref{Qeq}) reads 
\begin{equation}\label{QeqindeltaQ}
\frac{d (\delta Q)}{d\lambda} = -\delta Q \:[\delta U (1- 4\delta Q) + 
\delta Q (1-\delta Q +3 \delta U \delta Q)\:] \ .
\end{equation}
Since $\delta Q \rightarrow 0$ for $\lambda\rightarrow \infty$, for every 
small $\varepsilon >0$ there
exists some $\lambda_\varepsilon$ such that
\begin{equation}
\frac{d (\delta Q)}{d\lambda} \leq -\delta Q \,[(1-\varepsilon) \delta U]
\end{equation}
for all $\lambda > \lambda_\varepsilon$.

Consider the integral $\int_{\lambda_0}^\infty \delta U \delta Q \,d\lambda$ 
appearing in~(\ref{someint}), 
where $\lambda_0$ is chosen to be greater than $\lambda_\varepsilon$, 
where $\varepsilon$ is small.
We obtain
\begin{equation}
\int\limits_{\lambda_0}^\infty \delta U \delta Q \,d\lambda = 
\int\limits_0^{\delta Q(\lambda_0)} \delta U \delta Q\: 
\left(-\frac{d(\delta Q)}{d\lambda}\right)^{-1} d(\delta Q)
\leq \int\limits_0^{\delta Q(\lambda_0)} (1-\varepsilon)^{-1} d(\delta Q) < \infty 
\:,
\end{equation}
i.e., the integral is finite.
\proofend

\begin{theorem}\label{maintheorem2}
Let $\phi(\calE)$ be such that $n(\omega) \leq 5 + 3 l$ for all small $\omega$, and define
$\omega_{\mathrm{crit}} := \sup \{\omega\:|\:n(\omega)\leq 5 +3 l\}$, and $w_c := V_R -V(0)$.
Then every associated regular 
solution with $\omega_c \leq \omega_{\mathrm{crit}}$ has finite mass and radius.
\end{theorem}

\proof
Recall that for a regular solution the $\alpha$-limit set must
be a point on $L_2$.
With this information, the statement 
of the theorem can be deduced from Proposition \ref{globalomega3} in the 
same way that Theorem \ref{maintheorem1} was deduced from Proposition 
\ref{globalomega2}.
\proofend

\section{Conclusions and outlook}
\label{outlook}

In this paper we have investigated stationary solutions of the Vlasov-Poisson
system. We have derived conditions on the distribution function 
guaranteeing that the resulting steady states have compact support
and thus finite total masses.
Theorem~\ref{maintheorem1} formulates 
a condition that is local in the sense that 
only the properties of the distribution function for
particle energies close to the cut-off energy are relevant;
the theorem is a generalization of a theorem proved in~\cite{rr00}.
The condition of Theorem~\ref{maintheorem1} is 
not only sufficient but also necessary%
\footnote{When we restrict the theorem to the class of asymptotically
  polytropic distribution functions,
  this follows in a straightforward manner
  from a local dynamical systems analysis; we refer
  to the paper~\cite{hu1} where analogous issues
  occured in the context of static perfect fluid solutions.
  Generalizations to other classes of $\phi(\calE)$ are possible, but
  we refrain from discussing them here.}
for the statement to hold.
The condition stated in Theorem~\ref{maintheorem2}
is not local; this criterion is based on the
properties of the distribution function for a finite range of particle energies.
Theorem~\ref{maintheorem2} thus opens access to a class of distribution functions 
that could not be treated previously.
The condition of the theorem is sufficient, but not necessary.

In the astrophysical literature there exists a large variety
of stationary spherically symmetric models built of
self-gravitating collisionless matter.
Many of these models are based on
distribution functions for which the theorems of this paper
are relevant.
As examples we mention the King models, the Wooley--Dickens models,
see~\cite{binney}, and the Kent--Gunn models, see~\cite{kentgunn}:
Theorem~\ref{maintheorem1} applies.

Distribution functions of the form 
$\phi(\calE) \propto (\exp \calE - 1 - \calE) L^{2 l}$
---the isotropic case is called the Wilson model~\cite{binney}---
are in general not covered by Theorem~\ref{maintheorem1}.
However, Theorem~\ref{maintheorem2} applies for a wide range of $l$,
since $n(\omega)\rightarrow 7/2$ as $\omega\rightarrow 0$.
Theorem~\ref{maintheorem2} ensures finiteness of the radii and masses of solutions
that satisfy $\omega_c \leq \omega_{\mathrm{crit}}$;
since $n(\omega)$ is an increasing function which diverges as $\omega\rightarrow\infty$,
$\omega_{\mathrm{crit}}$ is necessarily a finite number.
The theorem does not give information
about radii and masses of regular solutions with $\omega_c > \omega_{\mathrm{crit}}$.
A numerical investigation suggests the following behaviour:
when $l=0$ all regular solutions, including those 
with $\omega_c >\omega_{\mathrm{crit}}$, possess compact support, which
is in agreement with~\cite{wilson}.
However, for a certain range of values of $l$, the situation
is more complex: for almost every value $\omega_c  >\omega_{\mathrm{crit}}$
the associated regular solution has a finite radius,
but there exists a discrete set of values $\{\omega_{c,i}\:|\: i =1 \ldots n, \,1\leq n < \infty\}$ 
such that the associated solutions
extend to infinity.
Hence, in many cases, but not in general,
the failure of 
the condition $n(\omega) \leq 5 +3 l$ to hold for large $\omega$
entails the occurence of regular solutions that are infinitely extended.

The above examples of distribution functions 
are all asymptotically polytropic in the low $\omega$ regime;
this allows one to also include the boundary $\Omega = 0$ in the dynamical
systems state space and to obtain specific information
about the asymptotic properties of solutions.
When the models are asymptotically polytropic or
``asymptotically isothermal'' in the high $\omega$ regime,
it is likewise possible to include $\Omega =1$
(although the isothermal case requires a slight change of variables).
In this context, one might be able to derive 
additional theorems concerning mass-radius properties 
of solutions in analogy to what was done in
the perfect fluid case in~\cite{hu1}, e.g.,
it might be possible to establish mass-radius relationships.

In the isotropic case the static Vlasov-Poisson system coincides
with the static Euler-Poisson system which describes
equilibrium states of self-gravitating perfect fluid matter.
The collisionless matter model with density $\rho$ and
radial pressure $p_{\mathrm{rad}}$, given in~(\ref{density}) 
and~(\ref{pressure}), can thus be interpreted as a perfect fluid matter
model with density $\rho$ and pressure $p=p_{\mathrm{rad}}$ with
a barotropic equation of state $\rho(p)$ implicitly determined by the distribution
function $\phi$.
It is straightforward to construct theorems that are analogous to~\ref{maintheorem1},~\ref{maintheorem2}
for the perfect fluid case; such theorems cover more general equations
of state than the class of asymptotically equations of state discussed in~\cite{hu1}.

In the present paper we have treated a collisionless gas in 
Newtonian theory of gravity. However, the methods used here
are likely to be of relevance also in the general relativistic 
case; the relativistic theorems in~\cite{rr00} might thus
be generalized.

\begin{appendix}

\section{Assumptions and properties of $\phi(\calE)$, $g_m(\omega)$, and $n(\omega)$}
\label{diffassumptions}

In this section we investigate the regularity assumptions on $\phi$ 
that guarantee that $n(\omega)$ becomes a $\mathcal{C}^1$-function for 
$\omega >0$.

By setting $x = \calE/\omega$ in the definition~(\ref{g}) of $g_m(\omega)$, we 
obtain
\begin{equation}\label{gm}
g_m(\omega) = \omega^{m+1} \int\limits_0^1 \phi(\omega x) 
(1-x)^m dx \:\qquad\quad (m>-1)\:.
\end{equation}

First some identities are obtained under the assumption that $\phi$ is
smooth. Then rougher functions $\phi$ are treated by approximating them
with smooth ones and passing to the limit. If $\phi$ is smooth then
differentiating (\ref{gm}) implies
\begin{equation}\label{directdiff}
\frac{d}{d\omega} g_m(\omega) =
(m+1)\omega^m\int\limits_0^1 \phi(\omega x)(1-x)^m dx + \omega^{m+1}
\int\limits_0^1 x \phi^\prime(\omega x) (1-x)^m d x.
\end{equation}
Integrating by parts gives the identity
\begin{equation}\label{dgm}
\frac{d}{d\omega} g_m(\omega) =
\omega^m \phi(\omega x) x (1-x)^m \Big|_0^1 + m g_{m-1}(\omega) =
m g_{m-1}(\omega)\:, 
\end{equation}
for $m>0$. For $m=0$ the result
$d g_0(\omega)/d \omega = \phi(\omega)$ is obtained.

The case $-1 < m < 0$ is more subtle. Each of the integrals on the right hand
side of (\ref{directdiff}) can be split into the sum of an integral from
zero to $1-\epsilon$ and an integral from $1-\epsilon$ to one for any 
$\epsilon$ between zero and one. As $\epsilon\to 0$ the second term
in each of the integrals tends to zero. The first term in each of the
integrals can be treated as was done for the integral from zero to one 
above. As a result
\begin{align*}
\frac{d}{d\omega} g_m(\omega) & = \omega^m \lim\limits_{\epsilon\rightarrow 0} 
\left[
\phi(\omega x) x (1-x)^m \Big|_0^{1-\epsilon} + m \int\limits_0^{1-\epsilon} \phi(\omega x) (1-x)^{m-1} d x 
\right] \\
& = 
\omega^m \lim\limits_{\epsilon\rightarrow 0}
\left[ 
\epsilon^m \Big( \phi(\omega- \epsilon\omega) -\phi(\omega)\Big) + \epsilon^m \phi(\omega) +
m \int\limits_0^{1-\epsilon} \phi(\omega x) (1-x)^{m-1} d x 
\right] \\
& = 
\omega^m \lim\limits_{\epsilon\rightarrow 0}
\left[
\frac{\phi(\omega- \epsilon\omega) -\phi(\omega)}{\epsilon^{-m}} + 
\phi(\omega) - m \int\limits_0^{1-\epsilon} \Big(\phi(\omega) -\phi(\omega x)\Big) (1-x)^{m-1} d x 
\right] \\
& = 
\omega^m \phi(\omega) + 
\omega^m \lim\limits_{\epsilon\rightarrow 0}
\left[
\epsilon^\delta \,\frac{\phi(\omega- \epsilon\omega) -\phi(\omega)}{\epsilon^{\alpha}} -
m \int\limits_0^{1-\epsilon} \frac{\phi(\omega) -\phi(\omega x)}{(1-x)^\alpha} (1-x)^{-1+\delta} d x
\right]\:,
\end{align*}
where $\alpha$ is defined as $\alpha = -m + \delta$ for an arbitrary $\delta >0$ (but preferably small).

Now suppose that $\phi$ is a measurable function, which is bounded on compact 
subsets of $(0,\infty)$ and satisfies 
$\phi(\calE) \leq \mathrm{const}\: \calE^k$ for 
some $k > -1$ on a neighbourhood of $\calE = 0$.
Then ${\cal E}^{-k}\phi$ is bounded on compact sets. It is possible to 
approximate it on any compact subset of $[0,\infty)$ by a sequence of smooth
functions which are uniformly bounded and converge to it pointwise. By
the dominated convergence theorem $g_m(\omega)$ is continuous and the 
sequence of functions $g_m$ defined by the approximants converge 
pointwise to that defined by $\phi$. Since (\ref{dgm}) holds for each of
the approximants it also holds for $\phi$ itself. It can be concluded that
$g_m$ is ${\mathcal C}^1$. In the case $m=0$, if $\phi$ is continuous then $g_m$ is
${\mathcal C}^1$. Equation (\ref{directdiff}) can alternatively be written as
\begin{equation}\label{dgmalt}
\frac{d}{d\omega} g_m(\omega) =
\omega^{-1} \Big(\: (m+1) g_m(\omega) + \omega^{m+1}
\int\limits_0^1 (\omega x) \phi^\prime(\omega x) (1-x)^m d x \:\Big)\:.
\end{equation}
If $\phi$ is ${\mathcal C}^1$ then we can use (\ref{dgmalt}) for each of the 
approximants and pass to the limit to get the same relation for $\phi$.
Accordingly, the function $g_m$ becomes $\mathcal{C}^2$ if
the second term in~(\ref{dgmalt}) is continously differentiable.
We note that this term is of the form~(\ref{gm})
where $\phi(y)$ has been replaced by $y \phi^\prime(y)$.
Using this formal resemblance we can simply 
adapt the conditions on $\phi$ discussed above;
hence, $g_m(\omega) \in \mathcal{C}^2(0,\infty)$,
if $\calE \phi^\prime(\calE)$ is 
bounded on compact subsets of $(0,\infty)$ and
$\calE \phi^\prime(\calE) \leq \mathrm{const}\: \calE^{k}$ 
for some $k > -1$ on a neighbourhood of $\calE = 0$.

Now consider the case $m<0$. When we assume that there exists $\delta>0$ such 
that $\phi \in \mathcal{C}^\alpha$, $\alpha =-m +\delta$, i.e., such that $\phi$ is 
H\"older-continuous with index $\alpha$, then the
approximants for $\phi$ can be chosen to converge in the appropriate 
H\"older norm and we can use the identity for the approximants and pass to
the limit to get the corresponding relation for $\phi$. Then 
\begin{equation}
\frac{d}{d\omega} g_m(\omega) =
\omega^m \phi(\omega) - m \omega^m \int\limits_0^1 
\frac{\phi(\omega)-\phi(\omega x)}{(1-x)^\alpha} (1-x)^{-1+\delta} dx \:,
\end{equation}
and it follows that $g_m$ is $\mathcal{C}^1$.

\end{appendix}


\end{document}